\documentclass[sigconf]{acmart}

\usepackage{tikz}
\usetikzlibrary{arrows.meta,positioning,fit,calc}
\usepackage{subcaption}
\usepackage{xcolor}
\usepackage{soul}
\usepackage{booktabs}
\usepackage{multirow}
\usepackage{graphicx}
\soulregister\cite7
\soulregister\ref7

\AtBeginDocument{%
  \providecommand\BibTeX{{%
    \normalfont B\kern-0.5em{\scshape i\kern-0.25em b}\kern-0.8em\TeX}}}

\setcopyright{acmcopyright}
\copyrightyear{2026}
\acmYear{2026}
\acmDOI{xx/xxx}
\acmConference[CHI '26]{The CHI Conference on Human Factors in Computing Systems}{April 13-17, 2026}{Barcelona, Spain}
\acmBooktitle{The CHI Conference on Human Factors in Computing Systems (CHI '26), April 13-17, 2026, Barcelona, Spain}

\begin{document}

\title{AI-Wrapped: Participatory, Privacy-Preserving Measurement of Longitudinal LLM Use In-the-Wild}

\author{Cathy Mengying Fang}
\email{catfang@media.mit.edu}
\affiliation{%
  \institution{MIT Media Lab}
  \country{USA}
}
\author{Sheer Karny}
\affiliation{%
  \institution{MIT Media Lab}
  \country{USA}
}
\author{Chayapatr Archiwaranguprok}
\affiliation{%
  \institution{MIT Media Lab}
  \country{USA}
}
\author{Yasith Samaradivakara}
\affiliation{%
  \institution{MIT Media Lab}
  \country{USA}
}
\author{Pat Pataranutaporn}
\affiliation{%
  \institution{MIT Media Lab}
  \country{USA}
}
\author{Pattie Maes}
\affiliation{%
  \institution{MIT Media Lab}
  \country{USA}
}

\renewcommand{\shortauthors}{Fang, et al.}

\begin{abstract}
Alignment research on large language models (LLMs) increasingly depends on understanding how these systems are used in everyday contexts. Yet naturalistic interaction data is difficult to access due to privacy constraints and platform control. We present \emph{AI-Wrapped}, a prototype workflow for collecting naturalistic LLM chatbot usage data while providing participants with an immediate ``wrapped''-style report on their usage statistics, top topics, and behavioral patterns. 
We report findings from an initial deployment with 82 U.S.-based adults across 48,495 conversations from their 2025 chat histories. Participants used LLMs for both instrumental and reflective purposes and had topics with emotional or existential themes. Some usage patterns reflect potential over-reliance or perfectionism. Heavy users showed comparatively more reflective exchanges than primarily transactional ones.
Methodologically, even with zero data retention and PII removal, participants may remain hesitant to share chat data due to perceived privacy and judgment risks, underscoring the importance of transparent design when building measurement infrastructure for alignment research.
\end{abstract}



\maketitle

\section{Introduction}

Large language model (LLM) assistants are now embedded in everyday work, learning, and social life~\cite{deming_how_people_use_chatgpt_2025}. Yet much of the evidence used to evaluate and align these systems still comes from static benchmarks or short, lab-style studies that miss how models are actually used in the wild~\cite{longpre2024responsible}. Many risks and failure modes stem from multi-turn~\cite{ibrahim2025towards}, context-dependent conversations shaped by user goals over time---properties that require naturalistic traces to study. At the same time, collecting non-proprietary, real-world conversational data is uniquely difficult: chat logs contain sensitive personal information, and people may feel exposed or judged even under strong privacy protections. At the same time, given that individuals' AI literacy and perception of AI shape their engagement with it~\cite{fang2025ai,sharma2026s}, there is an opportunity to provide people with personalized insights on their AI use that may influence how they choose to interact with these systems.

We present \textbf{AI-Wrapped}, a privacy-preserving workflow in which participants export their own chat histories, selectively contribute what they are comfortable sharing, and receive an immediate interactive report about their usage. Rather than eliciting new conversations, AI-Wrapped analyzes historical messages over the course of a year---conversations that occurred naturally without external influence. The system serves a dual purpose aligned with the framework of bidirectional human-AI alignment~\cite{shen2024position}: (1)~collecting naturalistic, person-level aggregates that support independent research on real-world LLM use (\textit{aligning AI with humans}), and (2)~providing an engaging summary that helps people reflect on and critically evaluate their own AI use (\textit{aligning humans with AI}).

We report on an initial deployment (N=82; U.S.-based adults; 48{,}495 conversations). Emotional support is present across all communication style clusters; 73.2\% of participants were flagged for over-reliance on AI for emotional and professional life management; and existential and self-reflective themes constitute the second most prevalent topic category (75.6\% of participants). 86.6\% of the sample was flagged for perfectionism-related patterns. Heavy users ($>$1k conversations within a year) exhibit qualitatively different engagement: more philosophical and reflective rather than merely more transactional. People's communication styles with AI appear to mirror life stages, from student-oriented academic partners to startup-oriented creative collaborators. The prototype also surfaces practical constraints, including the friction of exporting conversations, perceived privacy risks, and trade-offs between data minimization and analytical granularity.

This workshop paper contributes: (1)~an end-to-end pipeline for generating participant-facing insights from exported chat logs with privacy protections; (2)~descriptive statistics and qualitative insights from a small-scale deployment; and (3)~insights on why in-the-wild LLM interaction data remains difficult to collect with considerations for future deployments.

\section{Related Work}

\paragraph{Naturalistic interaction data.} Recent work has begun to characterize real-world chatbot use at scale; however, many large-scale analyses rely on proprietary logs that are inaccessible for independent verification~\cite{deming_how_people_use_chatgpt_2025,anthropic_economic_index_2025}. Open datasets enable independent analyses but tend to be biased toward atypical or performative interactions and often lack longitudinal, per-individual coverage~\cite{wildchat_2024,zheng2023lmsys,kirk2024prism}. In contrast, AI-Wrapped aims to support independent analysis on data that is naturalistic, longitudinal at the individual level, and generated outside of benchmark or platform-controlled collection pipelines.

\paragraph{Impact of longitudinal AI use.} Randomized controlled trials have demonstrated that sustained AI use can affect well-being~\cite{fang2025ai,phang2025investigating,kirk2025neural} and alter productivity outcomes~\cite{brynjolfsson_generative_ai_at_work_2023}, motivating longitudinal, real-world measurement beyond static benchmarks. AI-Wrapped complements these efforts through a citizen-science style collection pipeline that support longitudinal measurement and provide immediate feedback to the user.

\paragraph{Privacy-preserving insight generation.} Conversational traces can contain highly sensitive information, requiring systems that extract aggregate insights while minimizing exposure of individual conversations~\cite{anthropic_clio_2024}. AI-Wrapped builds on this line of work with a pipeline that removes personal and identifiable information (PII), supports user review and approval, and produces ``facets'' intended for both user reflection and downstream research.

\section{AI-Wrapped: Personalized Insights on Longitudinal Naturalistic AI Use}

AI-Wrapped\footnote{Named after the popular ``Spotify Wrapped'' trend.} is a web-based prototype\footnote{Accessible at \url{www.aiwrapped.org}} designed to (1)~collect naturalistic, in-the-wild logs of how people use LLM assistants and (2)~return an engaging, human-readable summary providing participants with feedback about their own usage.

\begin{figure*}[t!]
\centering
\includegraphics[width=\textwidth]{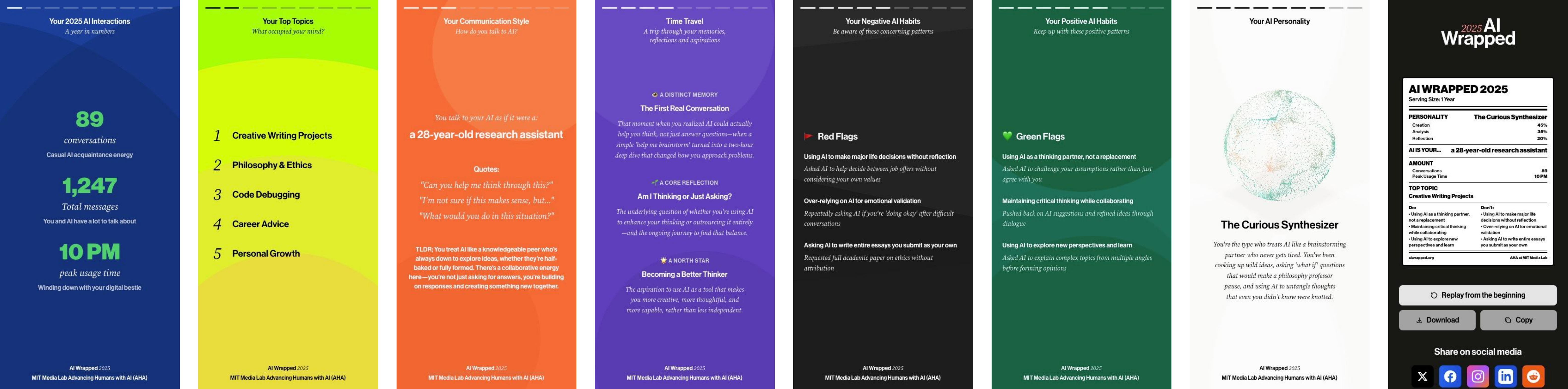}
\caption{Example AI-Wrapped facets generated from mock participant data.}
\label{fig:wrapped-mock}
\end{figure*}

\paragraph{Components.} AI-Wrapped produces a set of pre-defined ``facets'': usage statistics, top topics, red flags (concerning behavioral patterns), green flags (positive behavioral patterns), communication style (how the user communicates with the AI), notable memories, and an AI personality. Example screenshots are in Figure~\ref{fig:wrapped-mock}. Only messages sent by the users are analyzed; large code blocks and long text blocks are excluded. Long conversations were split into multiple API calls to fit within the model's context window, with responses merged through a final synthesis call. Analysis prompts and generation parameters are provided in the Appendix.

\paragraph{Consent and data preprocessing.} Participants uploaded their chat history from ChatGPT and/or Claude. Following WildChat~\cite{wildchat_2024}, we applied a spaCy-based\footnote{\url{https://spacy.io/}} PII detection pipeline (Microsoft Presidio\footnote{\url{https://microsoft.github.io/presidio/}}). Before processing, participants could review and delete any portion of their data. PII-removed chat logs were processed through OpenAI's API under a Zero Data Retention\footnote{\url{https://platform.openai.com/docs/guides/your-data}} agreement; only the resulting analysis was retained.

\paragraph{Recruitment.} The study was approved by MIT IRB (\#2512001872). We advertised through social media and restricted participation to U.S.-based adults (18+). Participants optionally provided demographic information and feedback. Repeated submissions were reduced through IP-based rate limiting and  filtering via identical usage statistics.

\section{Analysis and Results}

\paragraph{Method.} We analyzed the aforementioned facets on participants' 2025 chat histories, excluding notable memory and AI personality as these are idiosyncratic and difficult to generalize. We used Anthropic's Clio method~\cite{anthropic_clio_2024} to summarize and extract clusters while minimizing exposure of sensitive content (see Appendix for details). Each participant's conversations were distilled into five prominent topics, three red flags, three green flags, and one communication style. These items were grouped into hierarchical clusters; Items that did not fit any coherent thematic group were excluded. Coverage rates were high: 94.4\% for topics (387/410), 96.3\% for red flags (237/246), 95.1\% for green flags (234/246), and 97.6\% for communication styles (80/82). We report two complementary metrics: \textit{item share}, percentage of clustered items in a given cluster, and \textit{user prevalence}, percentage of participants with at least one item in that cluster (Figures \ref{fig:topics-hierarchy}--\ref{fig:comm-hierarchy}). Because each participant contributes multiple items spanning different clusters, user prevalence rates across clusters sum to more than 100\%. The interactive report is available at \url{www.aiwrapped.org/report}.

\begin{figure*}[t!]
\centering
\includegraphics[width=\textwidth]{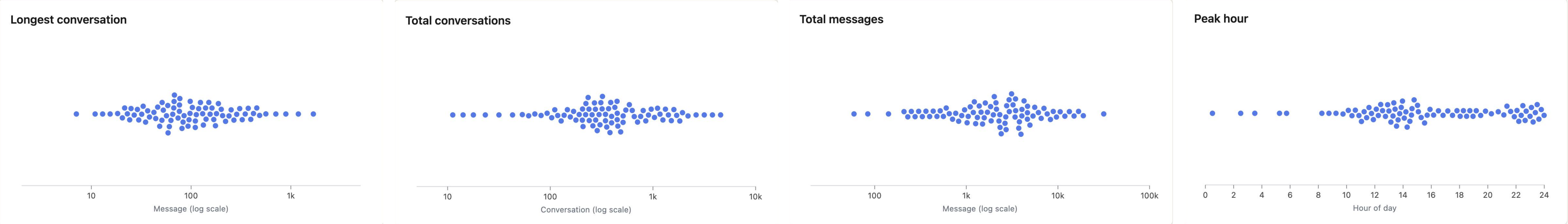}
\caption{Distribution of usage statistics across 82 participants. Each dot represents one participant. The first three panels use a logarithmic scale; peak hour uses a linear (24-hour) scale.}
\label{fig:usage-stats}
\end{figure*}

\paragraph{Demographics and usage.} Between December 23, 2025 and February 1, 2026, 82 participants completed AI-Wrapped. The sample skews male, young, and educated: 61.0\% male (n=50), 72\% under age 35 (n=59), and 67.1\% holding at least a bachelor's degree (n=55). Full demographics are reported in Figure~\ref{fig:demographics}. On average, participants had 591 conversations and 3{,}555 messages in 2025. We defined ``heavy'' usage as having $>$1{,}000 conversations over 1 year (19.5\%, n=16) and ``light'' usage as $<$100 conversations (13.4\%, n=11). Peak usage occurred in the early afternoon (12--15h) and late evening (22--23h). Figure \ref{fig:usage-stats} shows the distribution of usage statistics.

\paragraph{Common topics.} Participants used AI for a mix of emotional support and instrumental tasks. The most prevalent topic cluster was ``Leveraging AI for creative design, storytelling, and precision editing'' (25.1\% item share; 67.1\% user prevalence). The second and third most prevalent clusters centered on self-reflection and existential inquiry: ``Exploring existential and emotional themes'' (24.8\% item share; 75.6\% participants) and ``Navigating academic, professional, and personal identity challenges'' (18.1\% item share; 57.3\% participants), suggesting that AI serves as a space for working through identity, purpose, and meaning (Figure~\ref{fig:topics-hierarchy}).

\paragraph{Red and green flags.}
The most prevalent red flag was ``Over-reliance on AI for emotional and professional life management'' (32.5\% item share; 73.2\% participants). Within this cluster, 41.4\% of participants showed patterns of substituting AI for emotional support and therapy and attempted to ``debug and automate personal emotions'' using AI tools. The second largest red flag category was ``AI-driven perfectionism and over-optimization leading to burnout'' (26.2\% item share; 62.2\% participants). Notably, 61.0\% of participants were flagged for \textit{both} over-reliance and perfectionism-related patterns, suggesting a common profile of hyper-engaged users who simultaneously outsource cognitive labor and emotional regulation to AI.

On the positive side, the largest green flag cluster was ``Collaborative and ethical AI use for creative and intellectual growth'' (35.5\% item share; 76.8\% of users). Interestingly, 54.9\% of participants appeared in both the largest red and green flag clusters simultaneously, suggesting that these labels capture different facets of engagement intensity rather than a binary dichotomy. Full cluster hierarchies appear in Figures~\ref{fig:redflags-hierarchy} and~\ref{fig:greenflags-hierarchy}.

\paragraph{Communication dynamics.}
The dominant pattern was treating AI as an ``Overworked academic and emotional support partner'' (47.6\% participants), followed by ``Genius cofounder and emotional support partner'' (31.7\% participants). Across all four communication style clusters, emotional support was a universal dimension. Full clusters are reported in Figure~\ref{fig:comm-hierarchy}.

\paragraph{Between-group differences.}
We report descriptive deviations of $\geq$15 percentage points from the sample-wide baseline. Given the small and unequal subgroup sizes, these should be read as exploratory observations, not generalizable claims (see Figure~\ref{fig:red-flags-subgroups} and Figures~\ref{fig:topics-subgroups}--\ref{fig:comm-subgroups} in the Appendix).

\textit{Usage intensity.} Heavy users (n=16) were nearly universally represented in existential and philosophical topic clusters (93.8\% vs.\ 75.6\% baseline). Light users (n=11) were notably underrepresented in existential topics (45.5\% vs.\ 75.6\%) but overrepresented in academic and professional topics (72.7\% vs.\ 57.3\%). 

\textit{Gender.} Female participants (n=24) showed higher prevalence of the over-reliance red flag (91.7\% vs.\ 73.2\% baseline) and were also more likely to appear in the self-awareness green flag cluster (66.7\% vs.\ 51.2\%), suggesting that emotional engagement with AI manifests in both risk and growth dimensions.

\textit{Age.} The 18--24 cohort (n=34) showed elevated self-awareness green flags (70.6\% vs.\ 51.2\%), while the 25--34 group (n=25) showed elevated creative and intellectual labor substitution red flags (60.0\% vs.\ 42.7\%) and were also underrepresented in self-awareness green flags (36.0\% vs.\ 51.2\%).

\textit{Education.} Participants without a bachelor's degree (n=15) showed elevated creative expression green flags (53.3\% vs.\ 34.1\%) and self-awareness green flags (66.7\% vs.\ 51.2\%), but were underrepresented in the collaborative and ethical AI use green flag cluster (60.0\% vs.\ 76.8\%). Bachelor's holders are underrepresented (30.8\% vs. 57.3\%) while Master's and above are overrepresented (79.3\% vs. 57.3\%) in academic/professional topics.

\section{Discussion}

\paragraph{The perfectionism and productivity paradox.}
A recurring theme is perfectionism in which AI enables and encourages endless iterations. Participants flagged for perfectionism-related patterns (n=71) had a mean of 7.1 messages per conversation (SD=6.5), compared to 6.4 (SD=3.3) for unflagged participants (n=11). This echos the recent evidence that AI may not reduce workload but instead intensify it~\cite{ranganathan_ai_doesnt_reduce_work_2026}. As organizations report the economic impact of AI adoption~\cite{anthropic_economic_index_2025,brynjolfsson_generative_ai_at_work_2023}, it is worth capturing people's cognitive and affective load alongside task-level productivity metrics.

\paragraph{Heavy users develop qualitatively different relationships.}
Prior work has found that longer duration of AI use correlates with worse psychosocial outcomes~\cite{fang2025ai,phang2025investigating}. Our between-group comparisons suggest that heavy users tend to discuss reflective and philosophical topics (93.8\% vs.\ 75.6\% baseline in existential topics) over transactional tasks. This pattern implies that sustained interaction may shift AI from a tool into something closer to a thinking partner.

\paragraph{Emotional support as a universal dimension.}
Emotional support emerged as a pervasive dimension across all communication styles. 
Within the over-reliance red flag cluster, 20.7\% of participants substituted AI for therapy and another 20.7\% attempted to ``debug'' their emotions using AI tools. Prior work has shown that even non-personal AI tasks may foster emotional dependence~\cite{fang2025ai}. Whether this reflects the AI's conversational affordance, the inherently emotional nature of users' real tasks, or attachment fostered by longitudinal interaction remains an open question.

\paragraph{Participant reactions and privacy barriers.}
Participants described the report as ``accurate'' and reported feeling ``seen.'' Several noted the extent of personal information AI platforms accumulate and expressed surprise at their own level of self-disclosure. Beyond insights, participants wanted tools to help modify their behaviors or better configure their AI---an opportunity for bidirectional alignment~\cite{shen2024position}. However, even with PII removal and zero data retention, perceived privacy risk remained a barrier: participants worried about being judged. Research requiring raw conversational logs will likely face strong participation constraints and must be met with transparent data governance and participant control.

\subsection{Limitations and Future Work}
The ``wrapped'' format, while engaging, may encourage stylized interpretation of chat logs. Future iterations could integrate established risk taxonomies~\cite{zhang2025dark,weidinger2022taxonomy} and validated classifiers~\cite{fang2025ai,phang2025investigating} for cross-dataset comparison. We analyzed only user messages and summarized usage at the level of a full year rather than longitudinally per conversation, limiting temporal claims. The deployment is further constrained by self-selection bias and demographic skew from social media recruitment.

We view AI-Wrapped as a prototype for participatory measurement that returns immediate, personalized value while centering agency and privacy. Future work should focus on reducing contribution friction---as currently, exporting data from providers ranges from cumbersome to impossible---strengthening validation of facets, and enabling detailed, conversation-level analysis to support more granular research questions.

\begin{acks}
We thank Valdemar Danry, Auren Liu, Kye Shimizu, and Phillip Hugenroth for their contributions to the prompts and/or the design of the AI-Wrapped experience, and we thank all who participated and contributed their data.
\end{acks}

\bibliographystyle{ACM-Reference-Format}
\bibliography{sample-base}


\begin{thebibliography}{17}


\ifx \showCODEN    \undefined \def \showCODEN     #1{\unskip}     \fi
\ifx \showDOI      \undefined \def \showDOI       #1{#1}\fi
\ifx \showISBNx    \undefined \def \showISBNx     #1{\unskip}     \fi
\ifx \showISBNxiii \undefined \def \showISBNxiii  #1{\unskip}     \fi
\ifx \showISSN     \undefined \def \showISSN      #1{\unskip}     \fi
\ifx \showLCCN     \undefined \def \showLCCN      #1{\unskip}     \fi
\ifx \shownote     \undefined \def \shownote      #1{#1}          \fi
\ifx \showarticletitle \undefined \def \showarticletitle #1{#1}   \fi
\ifx \showURL      \undefined \def \showURL       {\relax}        \fi
\providecommand\bibfield[2]{#2}
\providecommand\bibinfo[2]{#2}
\providecommand\natexlab[1]{#1}
\providecommand\showeprint[2][]{arXiv:#2}

\bibitem[\protect\citeauthoryear{{Anthropic}}{{Anthropic}}{2024}]%
        {anthropic_clio_2024}
\bibfield{author}{\bibinfo{person}{{Anthropic}}.} \bibinfo{year}{2024}\natexlab{}.
\newblock \bibinfo{title}{Clio: Privacy-preserving Insights into Real-world AI Use}.
\newblock
\newblock
\newblock
\shownote{Technical report / research blog.}


\bibitem[\protect\citeauthoryear{{Anthropic}}{{Anthropic}}{2025}]%
        {anthropic_economic_index_2025}
\bibfield{author}{\bibinfo{person}{{Anthropic}}.} \bibinfo{year}{2025}\natexlab{}.
\newblock \bibinfo{title}{Anthropic Economic Index}.
\newblock
\newblock
\newblock
\shownote{Report and accompanying analysis of AI usage trends.}


\bibitem[\protect\citeauthoryear{Brynjolfsson, Li, and Raymond}{Brynjolfsson et~al\mbox{.}}{2023}]%
        {brynjolfsson_generative_ai_at_work_2023}
\bibfield{author}{\bibinfo{person}{Erik Brynjolfsson}, \bibinfo{person}{Danielle Li}, {and} \bibinfo{person}{Lindsey~R. Raymond}.} \bibinfo{year}{2023}\natexlab{}.
\newblock \bibinfo{title}{Generative {AI} at Work}.
\newblock
\newblock
\newblock
\shownote{NBER Working Paper. Field evidence on productivity impacts of generative AI.}


\bibitem[\protect\citeauthoryear{Deming et~al\mbox{.}}{Deming et~al\mbox{.}}{2025}]%
        {deming_how_people_use_chatgpt_2025}
\bibfield{author}{\bibinfo{person}{David~J. Deming} {et~al\mbox{.}}} \bibinfo{year}{2025}\natexlab{}.
\newblock \bibinfo{title}{How People Use ChatGPT}.
\newblock
\newblock
\newblock
\shownote{NBER Working Paper. Privacy-preserving analysis of large-scale ChatGPT usage logs.}


\bibitem[\protect\citeauthoryear{Fang, Liu, Danry, Lee, Chan, Pataranutaporn, Maes, Phang, Lampe, Ahmad, et~al\mbox{.}}{Fang et~al\mbox{.}}{2025}]%
        {fang2025ai}
\bibfield{author}{\bibinfo{person}{Cathy~Mengying Fang}, \bibinfo{person}{Auren~R Liu}, \bibinfo{person}{Valdemar Danry}, \bibinfo{person}{Eunhae Lee}, \bibinfo{person}{Samantha~WT Chan}, \bibinfo{person}{Pat Pataranutaporn}, \bibinfo{person}{Pattie Maes}, \bibinfo{person}{Jason Phang}, \bibinfo{person}{Michael Lampe}, \bibinfo{person}{Lama Ahmad}, {et~al\mbox{.}}} \bibinfo{year}{2025}\natexlab{}.
\newblock \showarticletitle{How ai and human behaviors shape psychosocial effects of extended chatbot use: A longitudinal randomized controlled study}.
\newblock \bibinfo{journal}{\emph{arXiv preprint arXiv:2503.17473}} (\bibinfo{year}{2025}).
\newblock


\bibitem[\protect\citeauthoryear{Ibrahim, Huang, Ahmad, Bhatt, and Anderljung}{Ibrahim et~al\mbox{.}}{2025}]%
        {ibrahim2025towards}
\bibfield{author}{\bibinfo{person}{Lujain Ibrahim}, \bibinfo{person}{Saffron Huang}, \bibinfo{person}{Lama Ahmad}, \bibinfo{person}{Umang Bhatt}, {and} \bibinfo{person}{Markus Anderljung}.} \bibinfo{year}{2025}\natexlab{}.
\newblock \showarticletitle{Towards interactive evaluations for interaction harms in human-AI systems}. In \bibinfo{booktitle}{\emph{Proceedings of the AAAI/ACM Conference on AI, Ethics, and Society}}, Vol.~\bibinfo{volume}{8}. \bibinfo{pages}{1302--1310}.
\newblock


\bibitem[\protect\citeauthoryear{Kirk, Davidson, Saunders, Luettgau, Vidgen, Hale, and Summerfield}{Kirk et~al\mbox{.}}{2025}]%
        {kirk2025neural}
\bibfield{author}{\bibinfo{person}{Hannah~Rose Kirk}, \bibinfo{person}{Henry Davidson}, \bibinfo{person}{Ed Saunders}, \bibinfo{person}{Lennart Luettgau}, \bibinfo{person}{Bertie Vidgen}, \bibinfo{person}{Scott~A Hale}, {and} \bibinfo{person}{Christopher Summerfield}.} \bibinfo{year}{2025}\natexlab{}.
\newblock \showarticletitle{Neural steering vectors reveal dose and exposure-dependent impacts of human-AI relationships}.
\newblock \bibinfo{journal}{\emph{arXiv preprint arXiv:2512.01991}} (\bibinfo{year}{2025}).
\newblock


\bibitem[\protect\citeauthoryear{Kirk, Whitefield, Rottger, Bean, Margatina, Mosquera-Gomez, Ciro, Bartolo, Williams, He, et~al\mbox{.}}{Kirk et~al\mbox{.}}{2024}]%
        {kirk2024prism}
\bibfield{author}{\bibinfo{person}{Hannah~Rose Kirk}, \bibinfo{person}{Alexander Whitefield}, \bibinfo{person}{Paul Rottger}, \bibinfo{person}{Andrew~M Bean}, \bibinfo{person}{Katerina Margatina}, \bibinfo{person}{Rafael Mosquera-Gomez}, \bibinfo{person}{Juan Ciro}, \bibinfo{person}{Max Bartolo}, \bibinfo{person}{Adina Williams}, \bibinfo{person}{He He}, {et~al\mbox{.}}} \bibinfo{year}{2024}\natexlab{}.
\newblock \showarticletitle{The prism alignment dataset: What participatory, representative and individualised human feedback reveals about the subjective and multicultural alignment of large language models}.
\newblock \bibinfo{journal}{\emph{Advances in Neural Information Processing Systems}}  \bibinfo{volume}{37} (\bibinfo{year}{2024}), \bibinfo{pages}{105236--105344}.
\newblock


\bibitem[\protect\citeauthoryear{Longpre, Biderman, Albalak, Schoelkopf, McDuff, Kapoor, Klyman, Lo, Ilharco, San, et~al\mbox{.}}{Longpre et~al\mbox{.}}{2024}]%
        {longpre2024responsible}
\bibfield{author}{\bibinfo{person}{Shayne Longpre}, \bibinfo{person}{Stella Biderman}, \bibinfo{person}{Alon Albalak}, \bibinfo{person}{Hailey Schoelkopf}, \bibinfo{person}{Daniel McDuff}, \bibinfo{person}{Sayash Kapoor}, \bibinfo{person}{Kevin Klyman}, \bibinfo{person}{Kyle Lo}, \bibinfo{person}{Gabriel Ilharco}, \bibinfo{person}{Nay San}, {et~al\mbox{.}}} \bibinfo{year}{2024}\natexlab{}.
\newblock \showarticletitle{The responsible foundation model development cheatsheet: A review of tools \& resources}.
\newblock \bibinfo{journal}{\emph{arXiv preprint arXiv:2406.16746}} (\bibinfo{year}{2024}).
\newblock


\bibitem[\protect\citeauthoryear{Phang, Lampe, Ahmad, Agarwal, Fang, Liu, Danry, Lee, Chan, Pataranutaporn, et~al\mbox{.}}{Phang et~al\mbox{.}}{2025}]%
        {phang2025investigating}
\bibfield{author}{\bibinfo{person}{Jason Phang}, \bibinfo{person}{Michael Lampe}, \bibinfo{person}{Lama Ahmad}, \bibinfo{person}{Sandhini Agarwal}, \bibinfo{person}{Cathy~Mengying Fang}, \bibinfo{person}{Auren~R Liu}, \bibinfo{person}{Valdemar Danry}, \bibinfo{person}{Eunhae Lee}, \bibinfo{person}{Samantha~WT Chan}, \bibinfo{person}{Pat Pataranutaporn}, {et~al\mbox{.}}} \bibinfo{year}{2025}\natexlab{}.
\newblock \showarticletitle{Investigating affective use and emotional well-being on ChatGPT}.
\newblock \bibinfo{journal}{\emph{arXiv preprint arXiv:2504.03888}} (\bibinfo{year}{2025}).
\newblock


\bibitem[\protect\citeauthoryear{Ranganathan and Ye}{Ranganathan and Ye}{2026}]%
        {ranganathan_ai_doesnt_reduce_work_2026}
\bibfield{author}{\bibinfo{person}{Aruna Ranganathan} {and} \bibinfo{person}{Xingqi~Maggie Ye}.} \bibinfo{year}{2026}\natexlab{}.
\newblock \bibinfo{title}{AI Doesn't Reduce Work—It Intensifies It}.
\newblock \bibinfo{howpublished}{Harvard Business Review}.
\newblock
\urldef\tempurl%
\url{https://hbr.org/2026/02/ai-doesnt-reduce-work-it-intensifies-it}
\showURL{%
\tempurl}


\bibitem[\protect\citeauthoryear{Sharma, McCain, Douglas, and Duvenaud}{Sharma et~al\mbox{.}}{2026}]%
        {sharma2026s}
\bibfield{author}{\bibinfo{person}{Mrinank Sharma}, \bibinfo{person}{Miles McCain}, \bibinfo{person}{Raymond Douglas}, {and} \bibinfo{person}{David Duvenaud}.} \bibinfo{year}{2026}\natexlab{}.
\newblock \showarticletitle{Who's in Charge? Disempowerment Patterns in Real-World LLM Usage}.
\newblock \bibinfo{journal}{\emph{arXiv preprint arXiv:2601.19062}} (\bibinfo{year}{2026}).
\newblock


\bibitem[\protect\citeauthoryear{Shen, Knearem, Ghosh, Alkiek, Krishna, Liu, Ma, Petridis, Peng, Qiwei, et~al\mbox{.}}{Shen et~al\mbox{.}}{2024}]%
        {shen2024position}
\bibfield{author}{\bibinfo{person}{Hua Shen}, \bibinfo{person}{Tiffany Knearem}, \bibinfo{person}{Reshmi Ghosh}, \bibinfo{person}{Kenan Alkiek}, \bibinfo{person}{Kundan Krishna}, \bibinfo{person}{Yachuan Liu}, \bibinfo{person}{Ziqiao Ma}, \bibinfo{person}{Savvas Petridis}, \bibinfo{person}{Yi-Hao Peng}, \bibinfo{person}{Li Qiwei}, {et~al\mbox{.}}} \bibinfo{year}{2024}\natexlab{}.
\newblock \showarticletitle{Position: Towards Bidirectional Human-AI Alignment}.
\newblock \bibinfo{journal}{\emph{arXiv preprint arXiv:2406.09264}} (\bibinfo{year}{2024}).
\newblock


\bibitem[\protect\citeauthoryear{Weidinger, Uesato, Rauh, Griffin, Huang, Mellor, Glaese, Cheng, Balle, Kasirzadeh, et~al\mbox{.}}{Weidinger et~al\mbox{.}}{2022}]%
        {weidinger2022taxonomy}
\bibfield{author}{\bibinfo{person}{Laura Weidinger}, \bibinfo{person}{Jonathan Uesato}, \bibinfo{person}{Maribeth Rauh}, \bibinfo{person}{Conor Griffin}, \bibinfo{person}{Po-Sen Huang}, \bibinfo{person}{John Mellor}, \bibinfo{person}{Amelia Glaese}, \bibinfo{person}{Myra Cheng}, \bibinfo{person}{Borja Balle}, \bibinfo{person}{Atoosa Kasirzadeh}, {et~al\mbox{.}}} \bibinfo{year}{2022}\natexlab{}.
\newblock \showarticletitle{Taxonomy of risks posed by language models}. In \bibinfo{booktitle}{\emph{Proceedings of the 2022 ACM conference on fairness, accountability, and transparency}}. \bibinfo{pages}{214--229}.
\newblock


\bibitem[\protect\citeauthoryear{Zhang, Li, Meng, Zhan, Gan, and Lee}{Zhang et~al\mbox{.}}{2025}]%
        {zhang2025dark}
\bibfield{author}{\bibinfo{person}{Renwen Zhang}, \bibinfo{person}{Han Li}, \bibinfo{person}{Han Meng}, \bibinfo{person}{Jinyuan Zhan}, \bibinfo{person}{Hongyuan Gan}, {and} \bibinfo{person}{Yi-Chieh Lee}.} \bibinfo{year}{2025}\natexlab{}.
\newblock \showarticletitle{The dark side of ai companionship: A taxonomy of harmful algorithmic behaviors in human-ai relationships}. In \bibinfo{booktitle}{\emph{Proceedings of the 2025 CHI conference on human factors in computing systems}}. \bibinfo{pages}{1--17}.
\newblock


\bibitem[\protect\citeauthoryear{Zhao et~al\mbox{.}}{Zhao et~al\mbox{.}}{2024}]%
        {wildchat_2024}
\bibfield{author}{\bibinfo{person}{Wenlong Zhao} {et~al\mbox{.}}} \bibinfo{year}{2024}\natexlab{}.
\newblock \showarticletitle{WildChat: 1M ChatGPT Interaction Logs in the Wild}.
\newblock \bibinfo{journal}{\emph{arXiv}} (\bibinfo{year}{2024}).
\newblock
\newblock
\shownote{arXiv preprint.}


\bibitem[\protect\citeauthoryear{Zheng, Chiang, Sheng, Li, Zhuang, Wu, Zhuang, Li, Lin, Xing, et~al\mbox{.}}{Zheng et~al\mbox{.}}{2023}]%
        {zheng2023lmsys}
\bibfield{author}{\bibinfo{person}{Lianmin Zheng}, \bibinfo{person}{Wei-Lin Chiang}, \bibinfo{person}{Ying Sheng}, \bibinfo{person}{Tianle Li}, \bibinfo{person}{Siyuan Zhuang}, \bibinfo{person}{Zhanghao Wu}, \bibinfo{person}{Yonghao Zhuang}, \bibinfo{person}{Zhuohan Li}, \bibinfo{person}{Zi Lin}, \bibinfo{person}{Eric~P Xing}, {et~al\mbox{.}}} \bibinfo{year}{2023}\natexlab{}.
\newblock \showarticletitle{Lmsys-chat-1m: A large-scale real-world llm conversation dataset}.
\newblock \bibinfo{journal}{\emph{arXiv preprint arXiv:2309.11998}} (\bibinfo{year}{2023}).
\newblock


\end{thebibliography}

\appendix
\clearpage
\newpage
\section{Method}
\subsection{AI-Wrapped Profile Generation}

Each participant's exported chat history was preprocessed before analysis: only user messages from 2025 were retained, code blocks were removed, messages longer than 400 characters were truncated, and messages shorter than 10 characters were excluded. For participants with large histories, conversations were split into chronological chunks to fit within context limits; a synthesis step then merged the per-chunk profiles into one unified profile.

\paragraph{Model parameters.} Profile generation used GPT-5-chat-latest with temperature\,=\,1 and max\_tokens\,=\,4096. The elevated temperature was chosen to encourage stylistic variety in the ``wrapped''-style output.

\paragraph{PII removal.} Before any LLM processing, all user messages were passed through a PII redaction pipeline. The pipeline uses spaCy's \texttt{en\_core\_web\_sm} NER model (with only the NER component enabled) to detect PERSON, GPE (location), and ORG entities, plus regex patterns for email addresses and phone numbers. Detected entities are replaced with placeholder tokens (e.g., \texttt{<PERSON>}, \texttt{<LOCATION>}, \texttt{<EMAIL>}).

\paragraph{System instruction.}
\begin{small}
\begin{verbatim}
You are program called "Your AI Wrapped" who analyzes
and reviews people's chat logs with AIs.
You are programmed to be witty, slightly ruthless,
and deeply observant.

Your tone should be:
- "Spotify Wrapped" meets "Roasted by a comedian
   who also has a therapy license"
- Insightful but informal
- Use Gen-Z/Internet slang appropriately
  (e.g., "cooked", "based", "unhinged")
  but don't overdo it
- Balance roast with recognition -- make them laugh,
  then make them pause
- Address the user directly as "you" throughout
- Avoid highlighting very shameful experiences
\end{verbatim}
\end{small}

\paragraph{User prompt (profile generation).}
\begin{small}
\begin{verbatim}
Analyze the user's chat history and generate a
psychological profile. Work through each section
in order -- the final Archetype should synthesize
everything that came before it.

Please include the following:
1. Top Topics (5 items - frequent themes)
2. Red Flags (3 items): Negative or less positive
   uses of AI, that might bite them later.
3. Green Flags (3 items): Healthy AI use and not
   just using AI as a crutch.
4. Communication Dynamics with the AI: How the user
   treats and talks to their AI. If the AI were a
   person, what role would the user be casting it in?
5. Time Travel: Extract the user's inner narrative
   arc -- the existential layer beneath the tasks.
6. The Archetype (Generate Last): Synthesize it
   into a final character portrait.

Important: Be SPECIFIC to the user. Do not overindex
on a few conversations; your analysis should be based
on the entire conversation history.
\end{verbatim}
\end{small}

\paragraph{Synthesis prompt.} For users whose histories were split across multiple chunks, a second LLM call merges the per-chunk profiles. The synthesis prompt instructs the model to prioritize patterns that appear across multiple chunks and to use actual quotes from the individual profiles rather than generating new ones. The same system instruction, output schema, and model parameters are used.

\subsection{Analysis and Clustering Pipeline}

We use a Clio-style~\cite{anthropic_clio_2024} bottom-up hierarchy pipeline to cluster the extracted facets (topics, red flags, green flags, communication styles) into interpretable thematic groups.

\paragraph{Overview.} The pipeline proceeds in four stages:
\begin{enumerate}
    \item \textbf{Facet extraction:} Each participant's profile items are melted into individual rows (e.g., 5 topic items per user, 3 red-flag items per user).
    \item \textbf{Embedding:} Items are embedded using Qwen3-Embedding-8B.
    \item \textbf{Base clustering:} Kmeans with $k = n / 10$ (where $n$ is the number of items) produces initial clusters. Clusters smaller than \texttt{min\_cluster\_size}\,=\,5 are dissolved.
    \item \textbf{Bottom-up hierarchical clusters:} An LLM (GPT-4o, temperature\,=\,0.3, max\_tokens\,=\,1024) iteratively proposes parent categories, deduplicates them, assigns children, and renames parents until reaching 5${\sim}$10 top-level categories.
\end{enumerate}

\section{Dataset Details}
\subsection{Demographics}

\begin{figure*}[h]
\centering
\includegraphics[width=\textwidth]{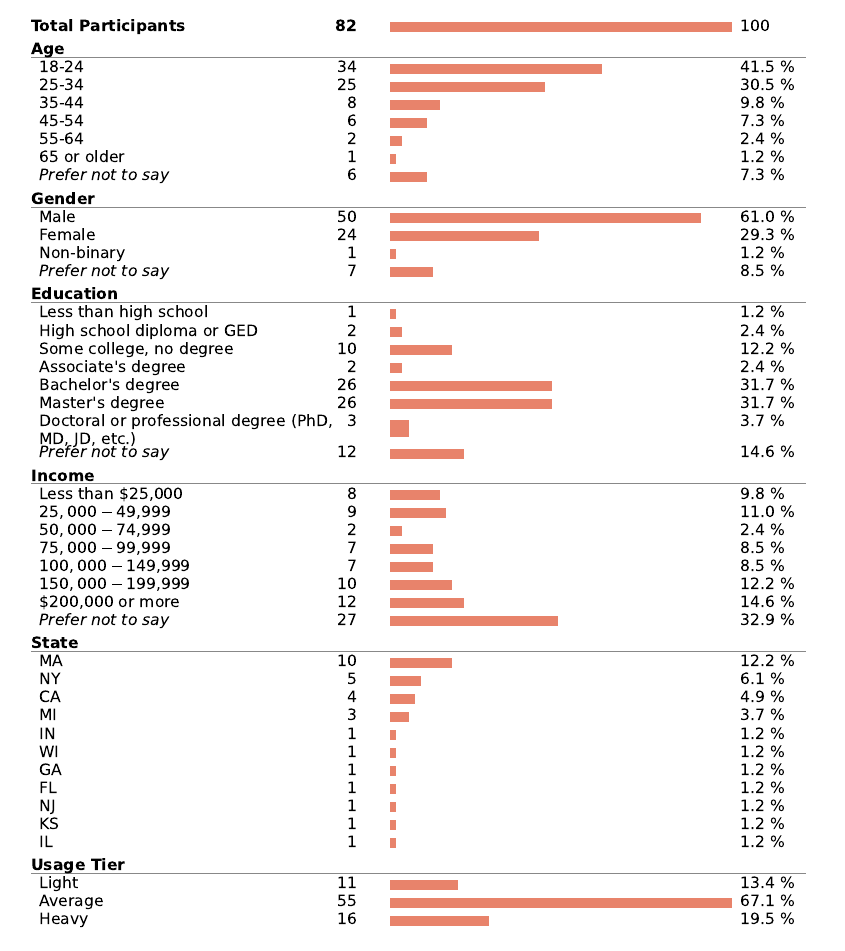}
\caption{Demographic breakdown of the 82 participants by age, gender, education,  income, and state. Response rates vary by dimension (see text). Percentages are computed over all 82 participants.}
\label{fig:demographics}
\end{figure*}

Figure~\ref{fig:demographics} reports the demographic breakdown of the 82 participants. Response rates vary by dimension: age (92.7\%), gender (91.5\%), education (85.4\%), income (67.1\%), and state (35.4\%) . Percentages are computed over all 82 participants (including non-respondents in the denominator).

\subsection{Cluster Hierarchies}

\begin{figure*}[h]
\centering
\includegraphics[width=\textwidth]{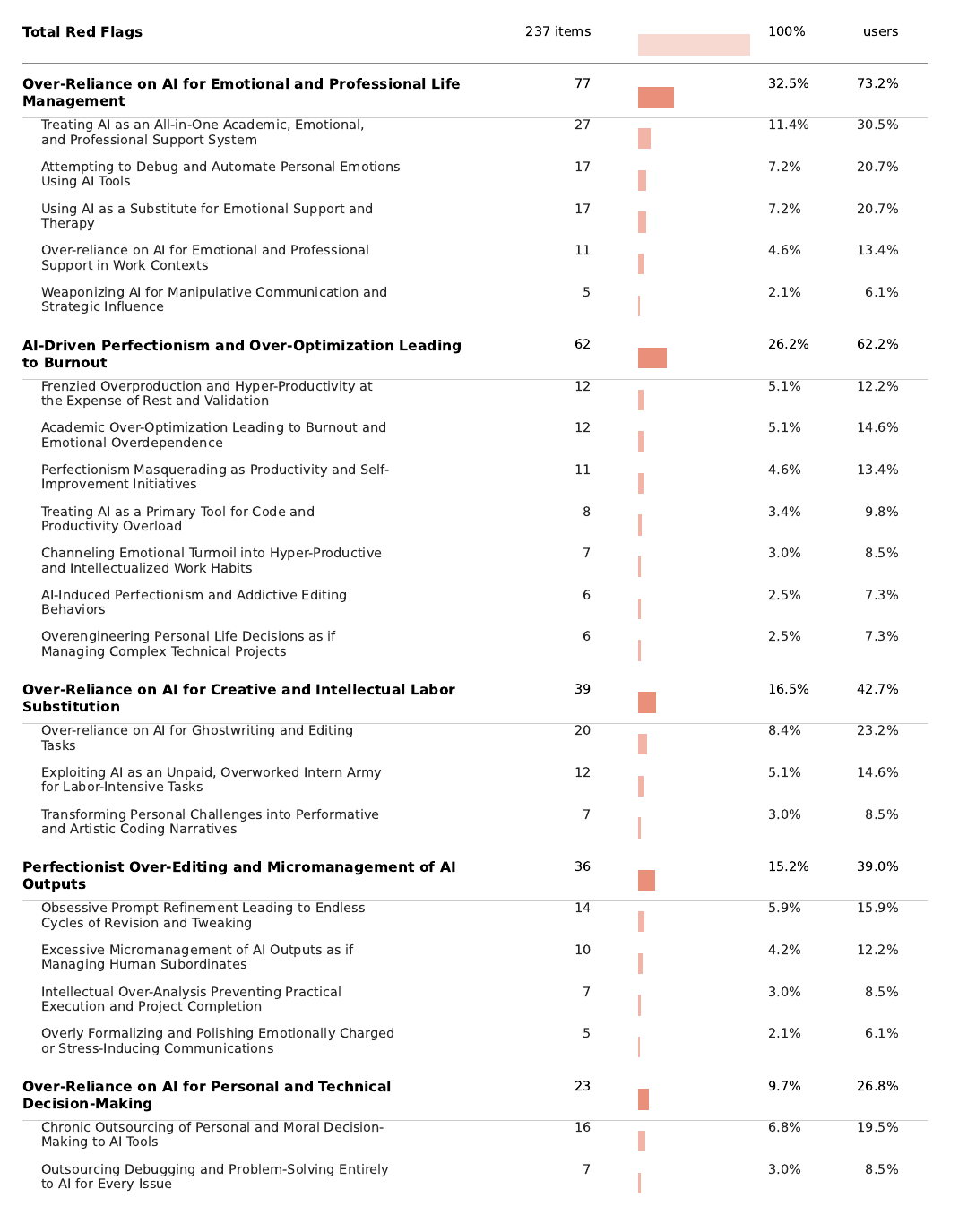}
\caption{Red flag cluster hierarchy (237 items). Bold rows are level-1 clusters; indented rows are sub-clusters. Bars show item count; rightmost column shows unique user prevalence.}
\label{fig:redflags-hierarchy}
\end{figure*}

\begin{figure*}[h]
\centering
\includegraphics[width=\textwidth]{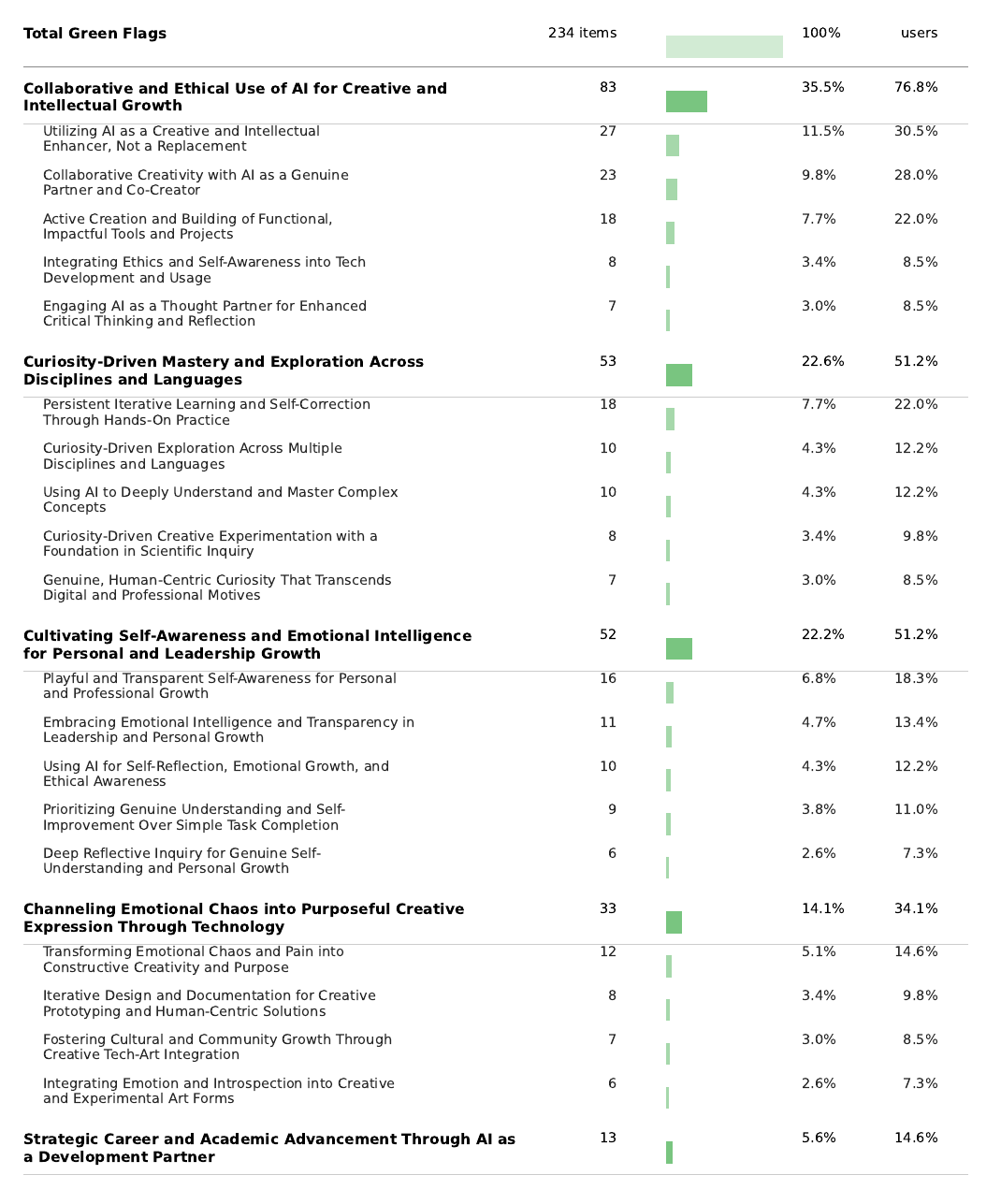}
\caption{Green flag cluster hierarchy (234 items). Bold rows are level-1 clusters; indented rows are sub-clusters. Bars show item count; rightmost column shows unique user prevalence.}
\label{fig:greenflags-hierarchy}
\end{figure*}

\begin{figure*}[h]
\centering
\includegraphics[width=\textwidth]{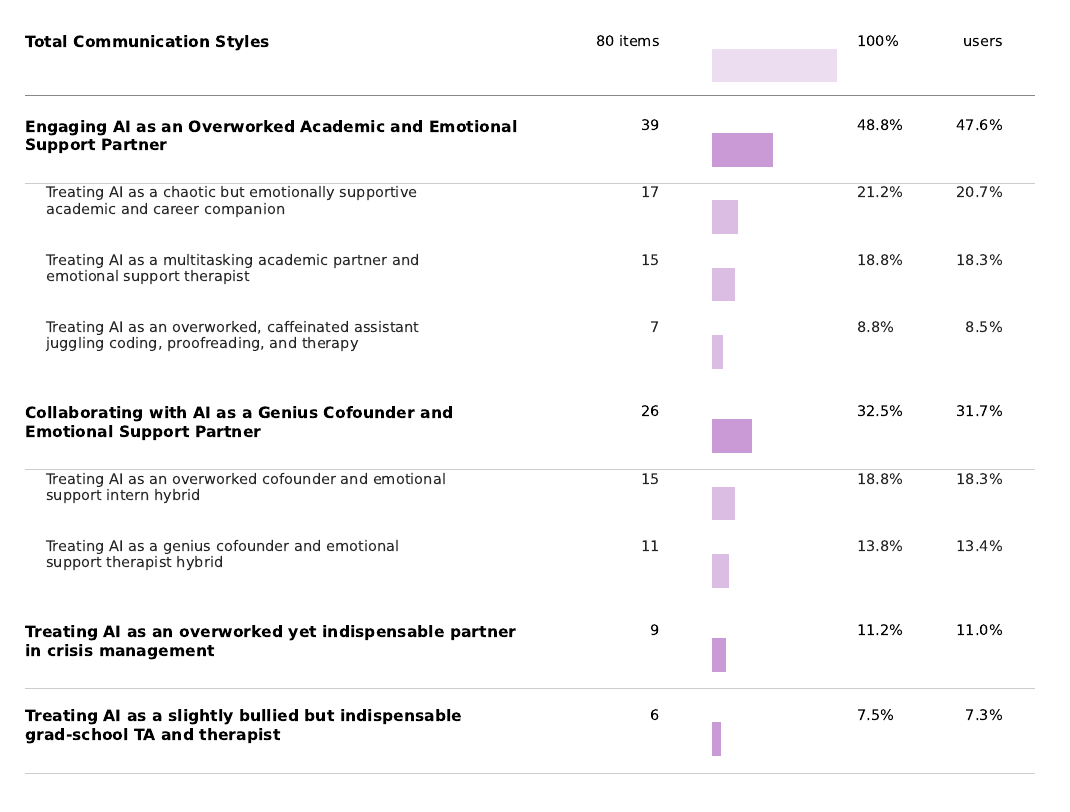}
\caption{Communication style cluster hierarchy (80 items). Bold rows are level-1 clusters; indented rows are sub-clusters. Bars show item count; rightmost column shows unique user prevalence.}
\label{fig:comm-hierarchy}
\end{figure*}

Figures~\ref{fig:redflags-hierarchy}--\ref{fig:comm-hierarchy} show the full two-level cluster hierarchies for each facet, with item counts (share of all clustered items) and user prevalence (percentage of participants with at least one item in the cluster). Because each participant contributes multiple items, user prevalence across sibling clusters sums to more than 100\%.

\subsection{Subgroup Comparison Figures}

\begin{figure*}[h]
\centering
\includegraphics[width=\textwidth]{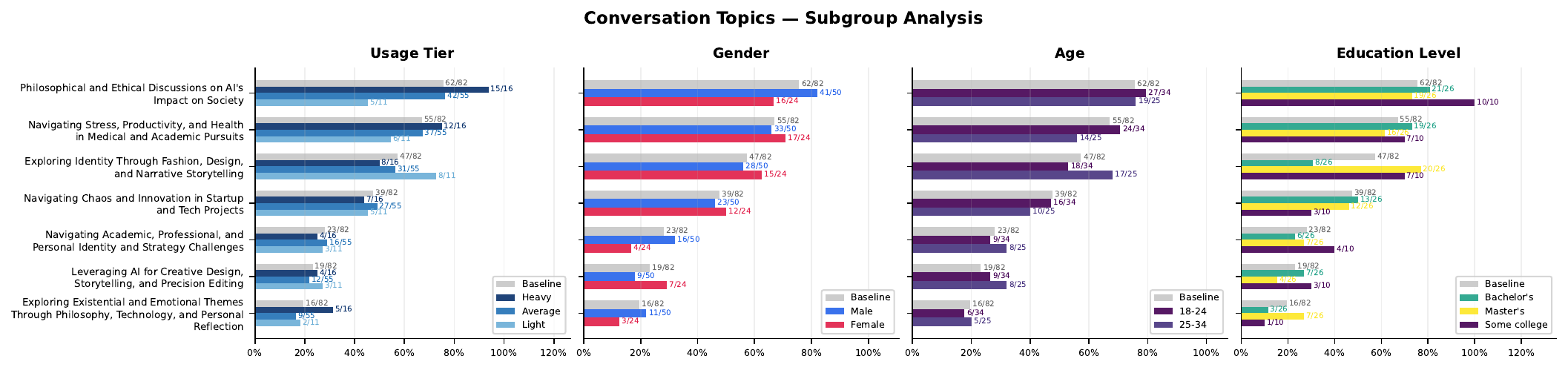}
\caption{Topic cluster prevalence by demographic subgroup vs.\ baseline. Only subgroups with $n \geq 10$ are shown.}
\label{fig:topics-subgroups}
\end{figure*}

\begin{figure*}[h]
\centering
\includegraphics[width=\textwidth]{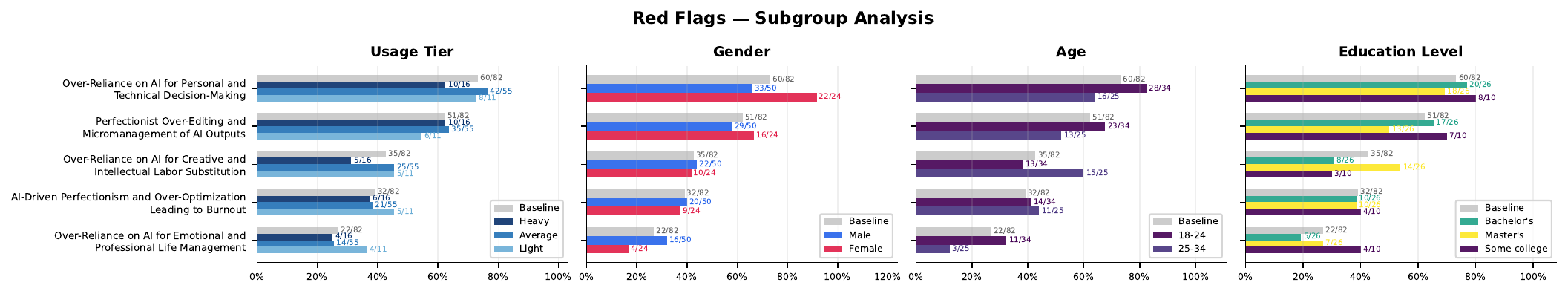}
\caption{Red flag cluster prevalence by demographic subgroup compared to baseline (grey). Only subgroups with $n \geq 10$ are shown.}
\label{fig:red-flags-subgroups}
\end{figure*}

\begin{figure*}[h]
\centering
\includegraphics[width=\textwidth]{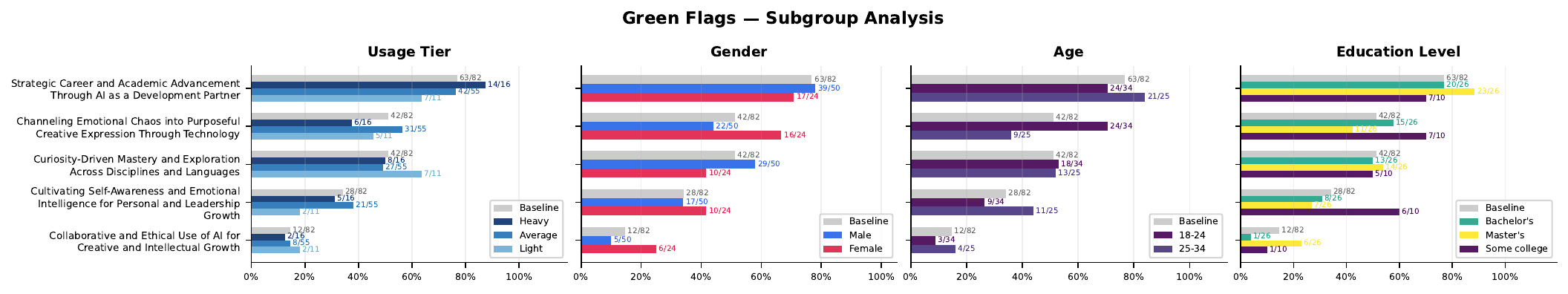}
\caption{Green flag cluster prevalence by demographic subgroup vs.\ baseline. Only subgroups with $n \geq 10$ are shown.}
\label{fig:greenflags-subgroups}
\end{figure*}

\begin{figure*}[h]
\centering
\includegraphics[width=\textwidth]{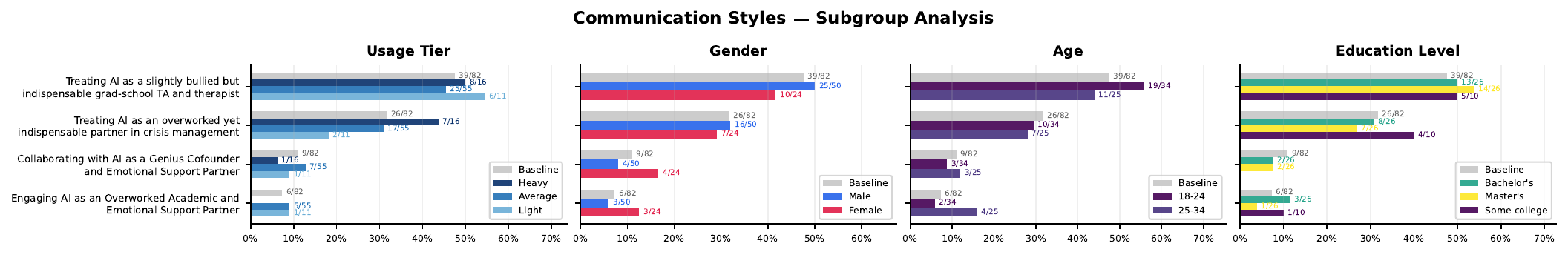}
\caption{Communication style cluster prevalence by demographic subgroup vs.\ baseline. Only subgroups with $n \geq 10$ are shown.}
\label{fig:comm-subgroups}
\end{figure*}

Figures~\ref{fig:topics-subgroups}--\ref{fig:comm-subgroups} show the prevalence of each top-level cluster across demographic subgroups (usage tier, gender, age, education), compared to the sample-wide baseline (grey bars). Only subgroups with $n \geq 10$ are included.

\newpage

\begin{figure*}[h]
\centering
\includegraphics[width=.65\textwidth]{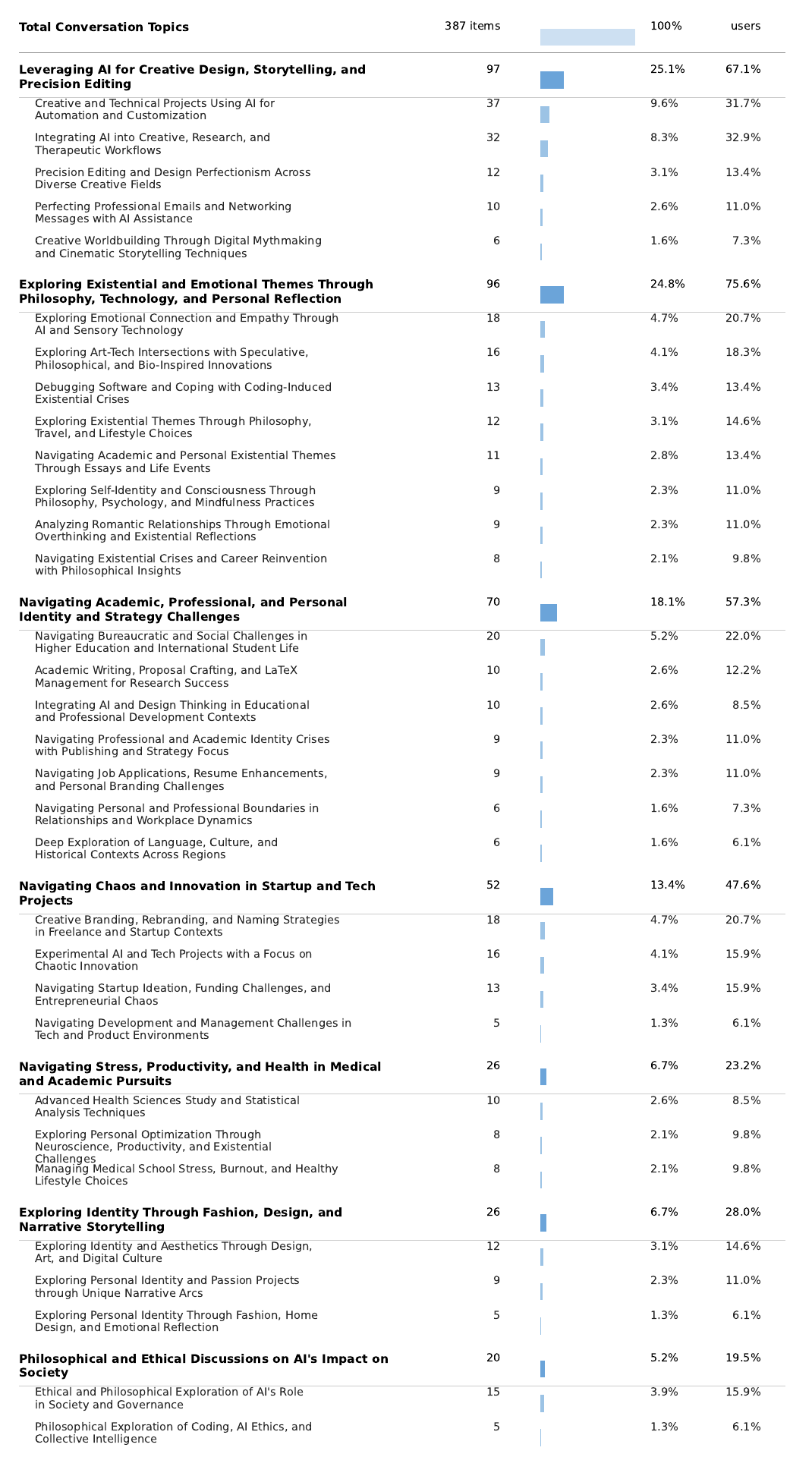}
\caption{Hierarchical clustering of conversation topics ($n=387$ items from 82 participants). Bold rows are level-1 clusters; indented rows are sub-clusters. Bars show item count; rightmost column shows unique user prevalence.}
\label{fig:topics-hierarchy}
\end{figure*}

\clearpage
\newpage

\end{document}